\documentclass{article}
\usepackage{spconf,amsmath,graphicx, hyperref}
\usepackage{xcolor}
\usepackage{tabularx}
\usepackage{makecell}
\usepackage{multirow}
\usepackage{booktabs}
\usepackage{subfig}
\usepackage[square, comma, sort&compress,numbers]{natbib}
\usepackage[accsupp]{axessibility} 
\usepackage{flushend,soul}

\newcommand\norm[1]{\left\lVert#1\right\rVert}

\title{ST-MFNet Mini: Knowledge Distillation-Driven Frame Interpolation}

\name{Crispian Morris, Duolikun Danier, Fan Zhang, Nantheera Anantrasirichai and David R. Bull \thanks{The work was funded by the University of Bristol, the UKRI MyWorld Strength in Places Programme (SIPF00006/1), and China Scholarship Council.}}
\address{Bristol Vision Institute, University of Bristol, One Cathedral Square, Bristol, BS1 5DD, UK.\\
{\{\href{mailto:crispian.morris@bristol.ac.uk}{crispian.morris}, \href{mailto:duolikun.danier@bristol.ac.uk}{duolikun.danier}, \href{mailto:fan.zhang@bristol.ac.uk}{fan.zhang}, \href{mailto:n.anantrasirichai@bristol.ac.uk}{n.anantrasirichai}, \href{mailto:dave.bull@bristol.ac.uk}{dave.bull}\}@bristol.ac.uk}
}

\begin{document}
\maketitle

\begin{abstract}

Currently, one of the major challenges in deep learning-based video frame interpolation (VFI) is the large model size and high computational complexity associated with many high performance VFI approaches. In this paper, we present a distillation-based two-stage workflow for obtaining compressed VFI models which perform competitively compared to the state of the art, but with significantly reduced model size and complexity. Specifically, an optimisation-based network pruning method is applied to a state of the art frame interpolation model, ST-MFNet, which suffers from large model size. The resulting network architecture achieves a 91\% reduction in parameter numbers and a 35\% increase in speed. The performance of the new network is further enhanced through a teacher-student knowledge distillation training process using a Laplacian distillation loss. The final low complexity model, ST-MFNet Mini, achieves a comparable performance to most existing high-complexity VFI methods, only outperformed by the original ST-MFNet. Our source code is available at \url{https://github.com/crispianm/ST-MFNet-Mini}

\end{abstract}


\section{Introduction}
\label{sec:intro}

Video frame interpolation (VFI) is a widely used technique for increasing the temporal resolution of video content. Recently, deep learning algorithms have enabled a significant boost in the performance of VFI methods~\cite{dong2022video}, and these have been employed extensively in various applications, including video compression, slow motion content rendering and view synthesis \cite{danier2022st}. 

Existing works on VFI generally rely on deep neural networks (DNNs) and can be classified as flow-based or kernel-based.  While flow-based VFI methods~\cite{jiang2018super, xu2019quadratic, niklaus2020softmax} estimate the optical flow to perform frame warping, kernel-based methods~\cite{niklaus2017video, lee2020adacof, choi2020channel} predict per-pixel interpolation kernels to synthesise the target frame. Recent advances in both classes include the use of coarse-to-fine architectures~\cite{reda2022film, kong2022ifrnet}, multi-stage/branch pipelines~\cite{bao2019depth, bao2019memc, danier2022st, gui2020featureflow}, 3D convolution~\cite{kalluri2020flavr, danier2022enhancing}, and self-attention mechanisms~\cite{lu2022video, shi2022video}. Although these latest developments have improved performance on commonly used benchmarks, there has been a tendency to adopt increasingly complex network designs, resulting in large model sizes and higher computational complexity. For example, the recently proposed model ST-MFNet~\cite{danier2022st} contains 21 million parameters, takes approximately 82MB to store in FP32 (32 bit floating point precision) and requires much longer runtime than much simpler VFI approaches (e.g., 45$\times$ compared to AdaCoF \cite{lee2020adacof}). Such a large model is inefficient in both training and inference (thus having high carbon footprint \cite{lacoste2019quantifying}), imposing high memory requirements and restricting deployment in real-world scenarios.

In order to tackle the problem of large model sizes and high complexity, techniques such as model pruning~\cite{reed1993pruning} and knowledge distillation~\cite{hinton2015distilling} are commonly utilised. One of the few efforts made to apply these to VFI is by Ding et al.~\cite{ding2021cdfi}, pruning  the AdaCoF model~\cite{lee2020adacof} using a sparsity-inducing optimisation objective~\cite{chen2021orthant} to obtain a small network with evidently decreased performance. Newly designed model components were then integrated into this small model to enhance its performance. It is noted that, although a large compression ratio was achieved by pruning the model, the additional components that contributed to the performance gain of the pruned model doubled its size.

\begin{figure}
    \centering
    \includegraphics[width=\linewidth]{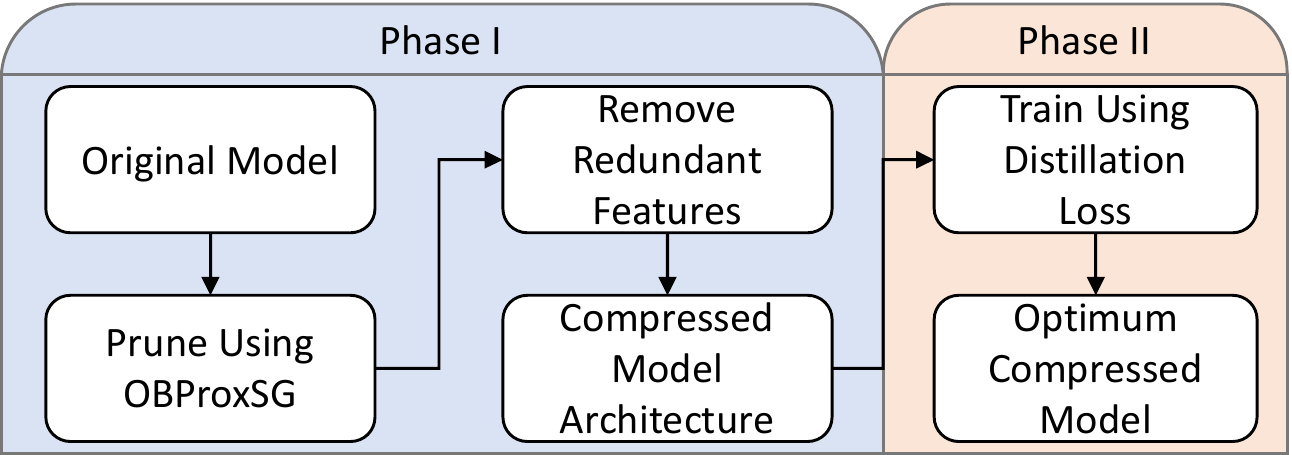}
    \caption{The proposed VFI compression workflow. Phase I: model size and complexity reduction. Phase II: interpolation performance enhancement.}
    \label{fig:workflow}
\end{figure}

In this work, we leverage knowledge distillation to achieve model compression for VFI. This allows a compact model to obtain favourable performance against the state of the arts, without including any additional modules. Specifically, we design a two-stage pipeline (illustrated in \autoref{fig:workflow}) to compress ST-MFNet - a recent state-of-the-art VFI model. In the first stage, we follow \cite{ding2021cdfi}, adopting a sparsity-induced model pruning technique, OBProx-SG~\cite{chen2021orthant}, to obtain a new, compact ST-MFNet with 91\% reduction in the number of parameters. In the second stage, we train using a novel knowledge distillation loss so that it `learns' additional information from the `teacher' model (a pre-trained original ST-MFNet). The resulting low complexity VFI approach, ST-MFNet Mini, outperforms its `teacher' in certain cases, with significantly reduced model size and complexity. To the best of our knowledge, this is the first attempt that employs knowledge distillation for the purpose of model compression in VFI. The proposed framework can potentially be applied to other VFI methods to reduce their model complexity.

\section{Proposed Method}
\label{sec:methods}

In this work, we configure the model as in \cite{danier2022st}; given four consecutive frames $\{I_0, I_1, I_2, I_3\}$, the VFI method outputs $I_t$ where $t=1.5$ to achieve $\times$2 interpolation. In this section, we first describe the ST-MFNet architecture and the process of pruning it (Phase I in \autoref{fig:workflow}), followed by the knowledge distillation process used to train the compact ST-MFNet (ST-MFNet Mini).

\subsection{Pruning ST-MFNet}

Given four input frames, the original ST-MFNet first processes these in two branches: MIFNet and BLFNet. While the former uses a U-Net style model to predict per-pixel deformable interpolation kernels, the latter employs a pre-trained optical flow estimator to obtain one-to-one pixel correspondence. These estimated kernels and optical flows are then used to synthesise the target middle frame. The results are passed to a GridNet~\cite{fourure2017residual}, from which we obtain an intermediate interpolation result, $\Tilde{I}_t$. This is combined with all four input frames and fed to a 3D CNN to estimate the textural residuals, denoted as $R$. The final output is the combination of these terms,  $\tilde{I}_t + R$. Readers are referred to~\cite{danier2022st} for further details.

\noindent\textbf{Sparsity inducing optimisation.} In order to obtain a condensed version of ST-MFNet, we start with the original pre-trained model, which is publicly available \href{https://drive.google.com/file/d/1s5JJdt5X69AO2E2uuaes17aPwlWIQagG/view?usp=sharing}{here}, and fine-tune it using the following loss function,
\begin{gather} 
    \mathcal{L}_\mathrm{Charb}(I_{pred}, I_{gt}) = \sqrt{(I_{pred}-I_{gt})^2 + \epsilon^2}, \\
    \mathcal{L}_\mathrm{prune} = \mathcal{L}_\mathrm{Charb}(I_{pred}, I_{gt}) + \lambda \norm{\theta}_1
\end{gather}
where $I_{gt}$ and $I_{pred}$ are the ground-truth target frame and the prediction of ST-MFNet respectively, and $\epsilon = 0.001$. The parameters of ST-MFNet are denoted as $\theta$, and $\norm{\theta}_1$ refers to the $l_1$ norm regularisation term, which induces sparsity in the network~\cite{chen2021orthant}. Such sparsity information can be used as a guide to remove redundant layers in the network. We adopt the OBProx-SG~\cite{chen2021orthant} solver to perform the optimisation, which has been shown to be efficient for sparsity-based model pruning~\cite{ding2021cdfi}. 

Similarly to~\cite{ding2021cdfi}, we define each layer, $l$, in the model as $l = [C_{out}, C_{in}, q_0, q_1, q_2]$, with total number of parameters $p_l = C_{out} \cdot C_{in} \cdot q_0 \cdot q_1 \cdot q_2$. As training takes place, we calculate the density ($d$) for each layer $l$ using 
\begin{equation}
    d_l = \frac{\text{\# of nonzero weights in } l}{p_l},
\label{eq:density}
\end{equation}
and use these values to compute the average density in the model as an indicator for the training progress. Following~\cite{ding2021cdfi}, we set $\lambda=10^{-4}$ and optimise the network for 20 epochs using our combined training set (Sec.~\ref{sec:results-setup}), achieving a density of approximately 0.24.

\textbf{Network compression}. The sparsity-inducing optimisation described above allows us to identify, through the final layer densities, the individual network layers which contribute less to the overall model performance, and thus can be removed or shrunk. Therefore, starting with the final layer, we use its density as its compression ratio ($r_n$), forming a new layer in the form 
\begin{equation*}
    l_n' = [C_{out}, r_n \cdot C_{in}, q_0, q_1, q_2],
\end{equation*}
and causing 
\begin{equation*}
    l_{n-1}' = [r_{n-1} \cdot C_{out}, C_{in}, q_0, q_1, q_2].
\end{equation*}
We iterate through the model's $n$ layers in this fashiopn, as in~\cite{chen2020neural}, ensuring that each layer's outputs match the previously compressed layer's inputs.

\begin{figure*}[t]
    \centering
    \includegraphics[width=1\linewidth]{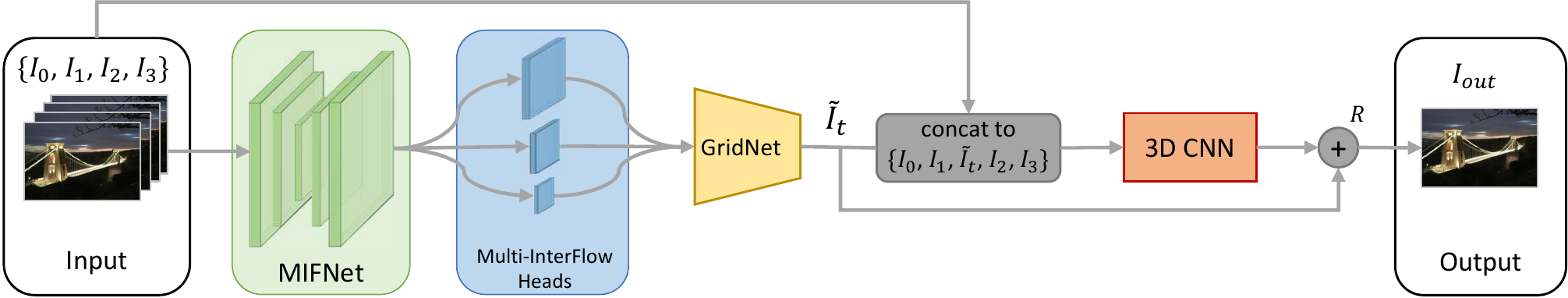}
    \caption{The resulting model architecture of ST-MFNet Mini.}
    \label{fig:architecture}
\end{figure*}

For ST-MFNet, we found that the flow estimator component of the model (BLFNet) has near-zero compression ratios, so it is removed entirely instead of shrunk. Ultimately, this process prunes the model from 21.03M parameters to 1.82M, a 91\% reduction. The architecture of the pruned network (ST-MFNet Mini) is shown in \autoref{fig:architecture}

\begin{figure*}[t]
    \centering

    \setcounter{subfigure}{0}
    \subfloat[Overlay] {\includegraphics[width=0.12\linewidth]{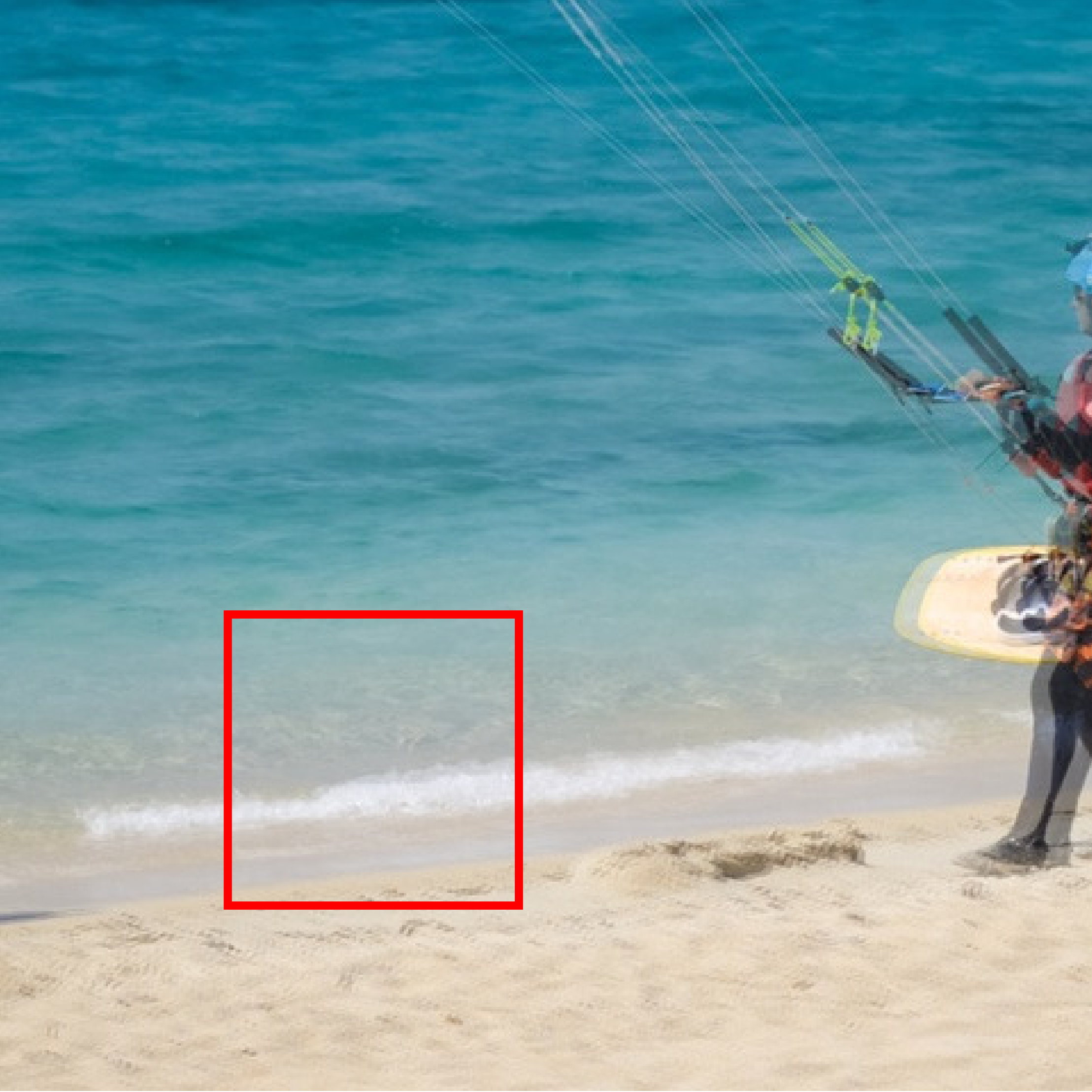}}\;\!\!
	\subfloat[ST-MFNet] {\includegraphics[width=0.12\linewidth]{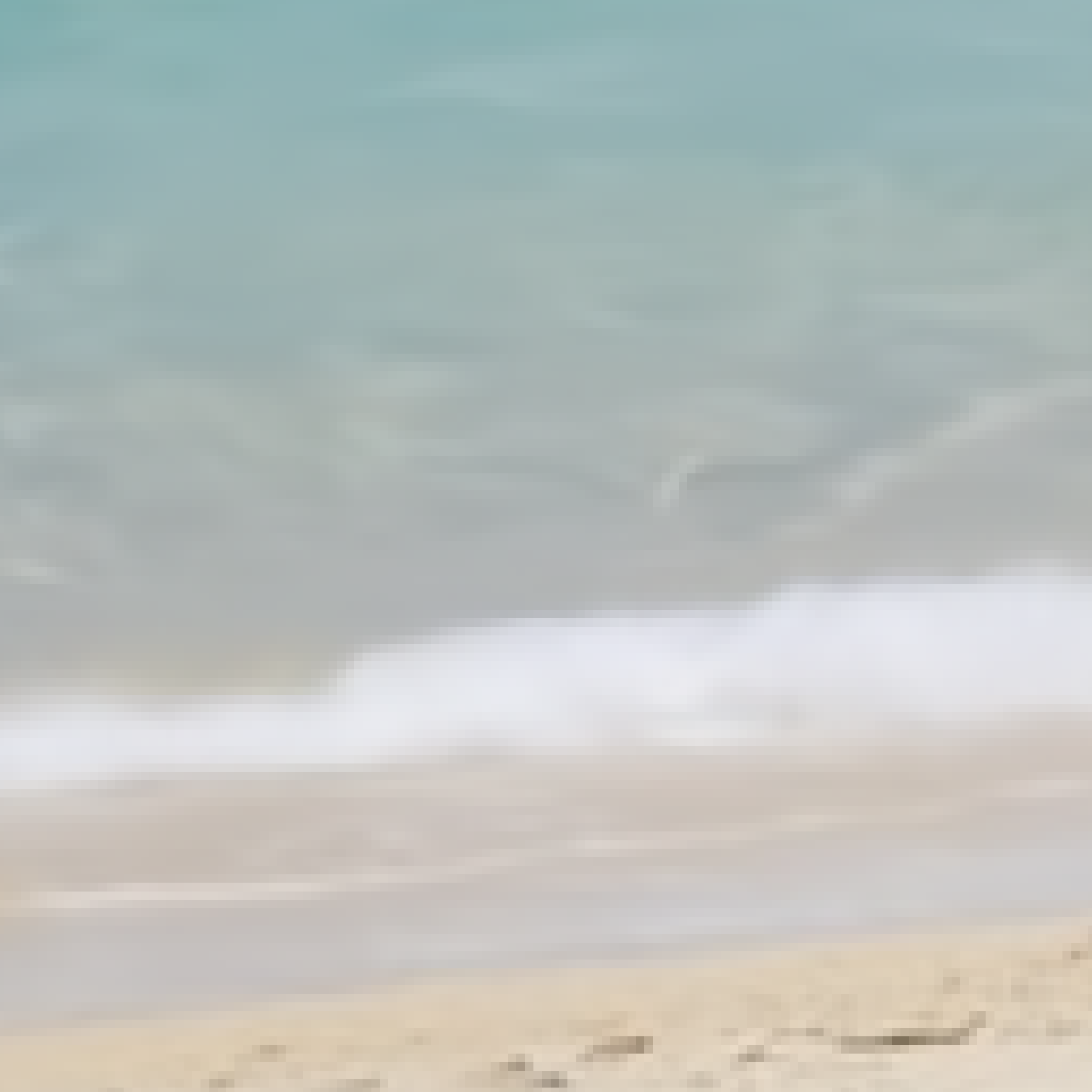}}\;\!\!
	\subfloat[Ours] {\includegraphics[width=0.12\linewidth]{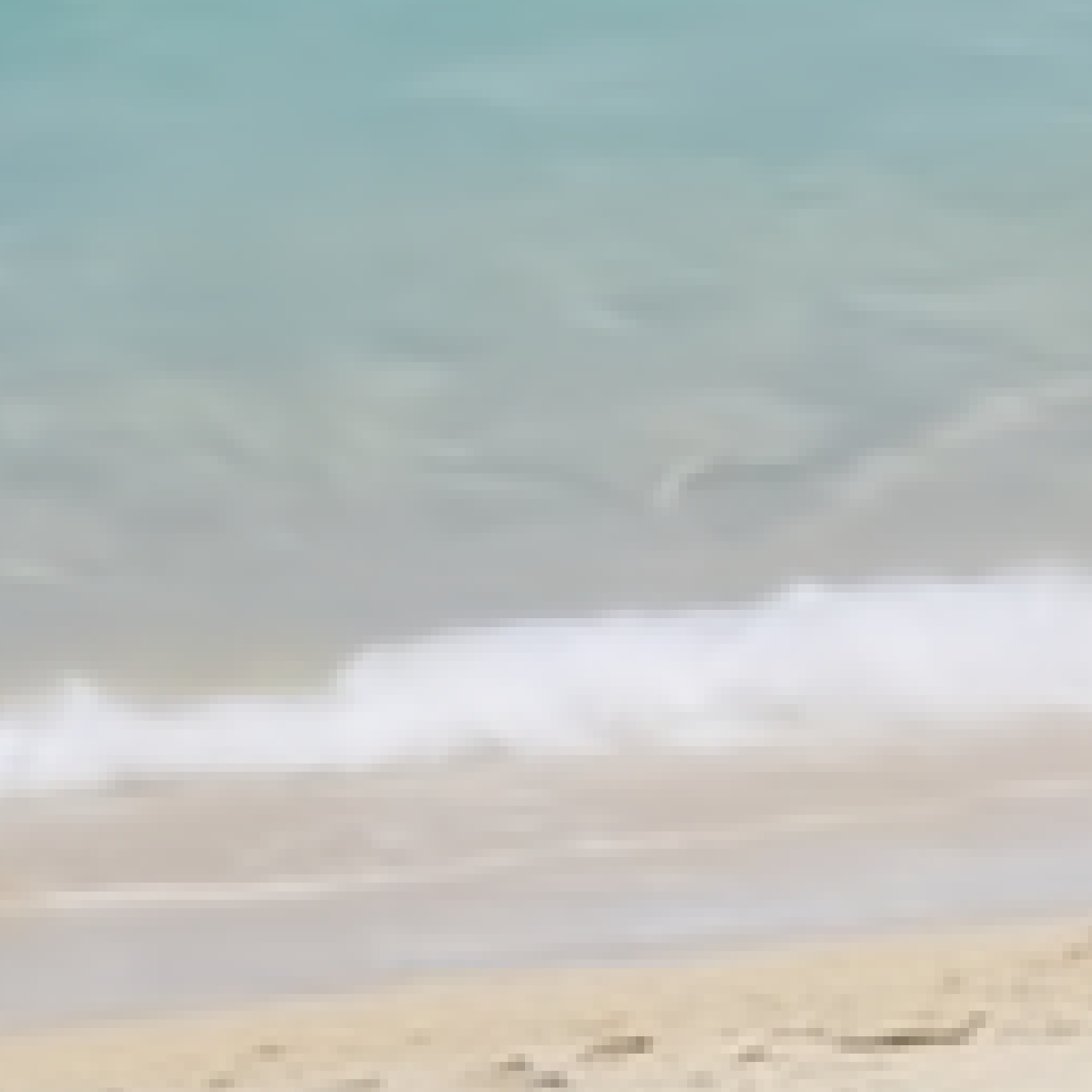}}\;\!\!
    \subfloat[GT] {\includegraphics[width=0.12\linewidth]{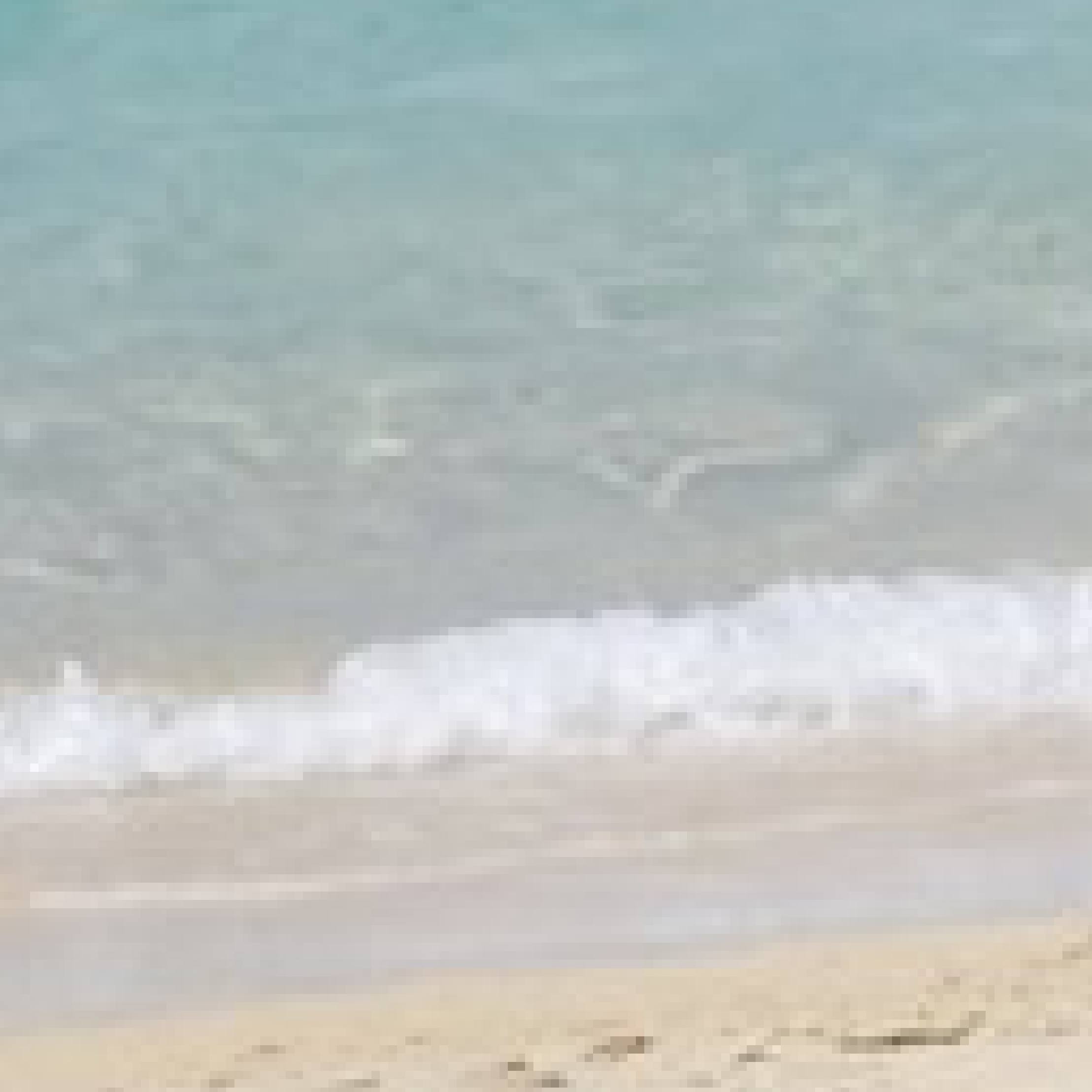}}\;\!\!
    \subfloat[Overlay] {\includegraphics[width=0.12\linewidth]{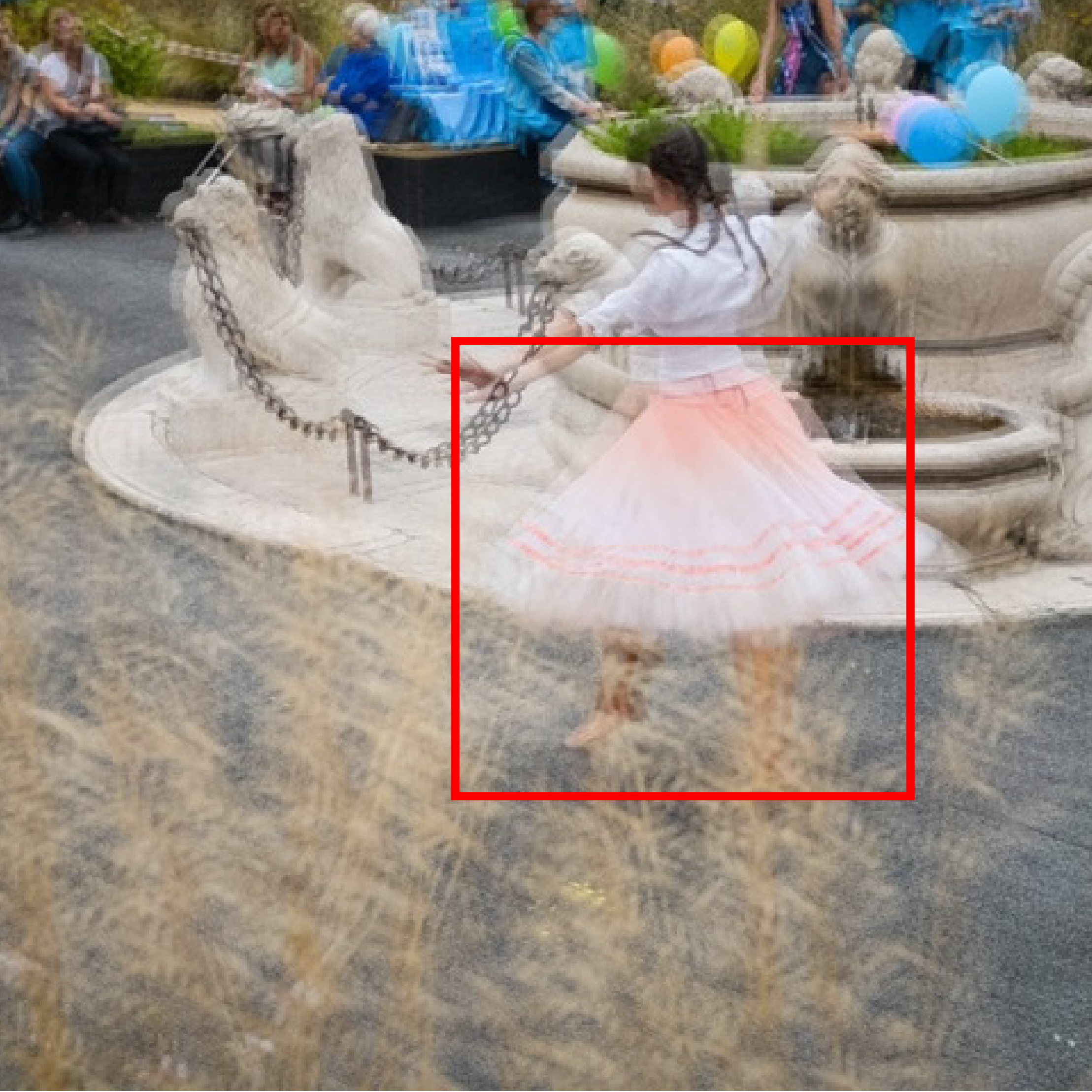}}\;\!\!
    \subfloat[ST-MFNet] {\includegraphics[width=0.12\linewidth]{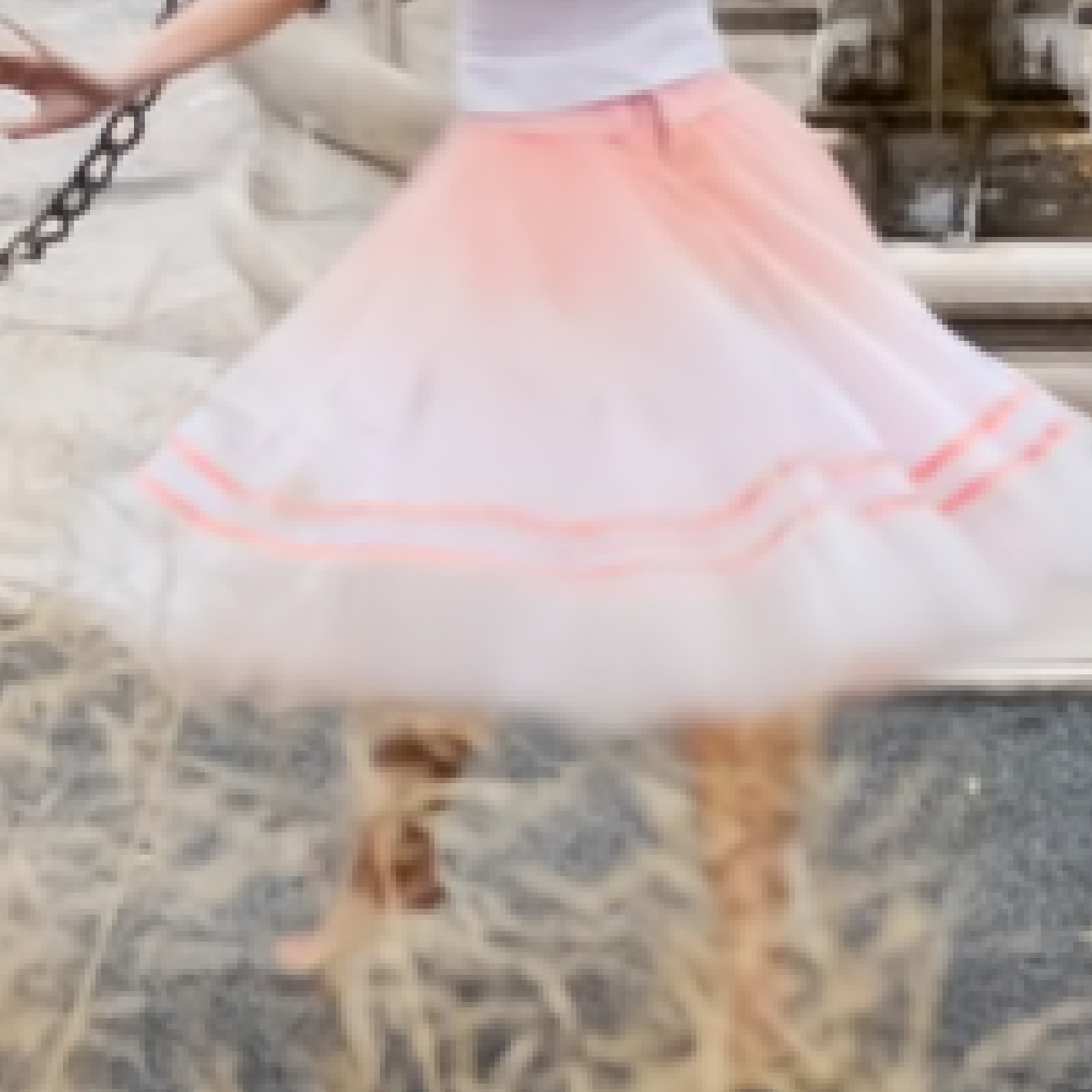}}\;\!\!
    \subfloat[Ours] {\includegraphics[width=0.12\linewidth]{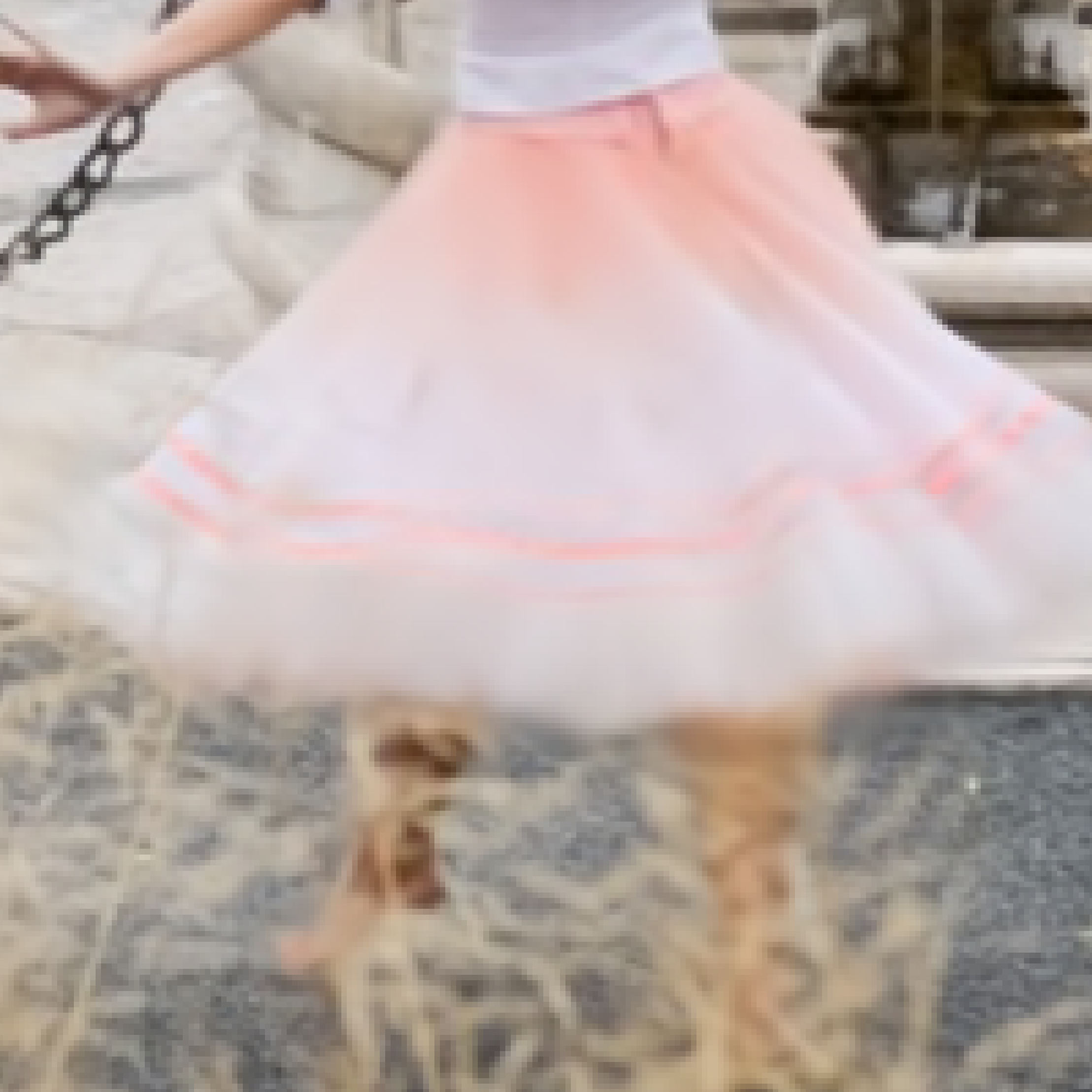}}\;\!\!
    \subfloat[GT] {\includegraphics[width=0.12\linewidth]{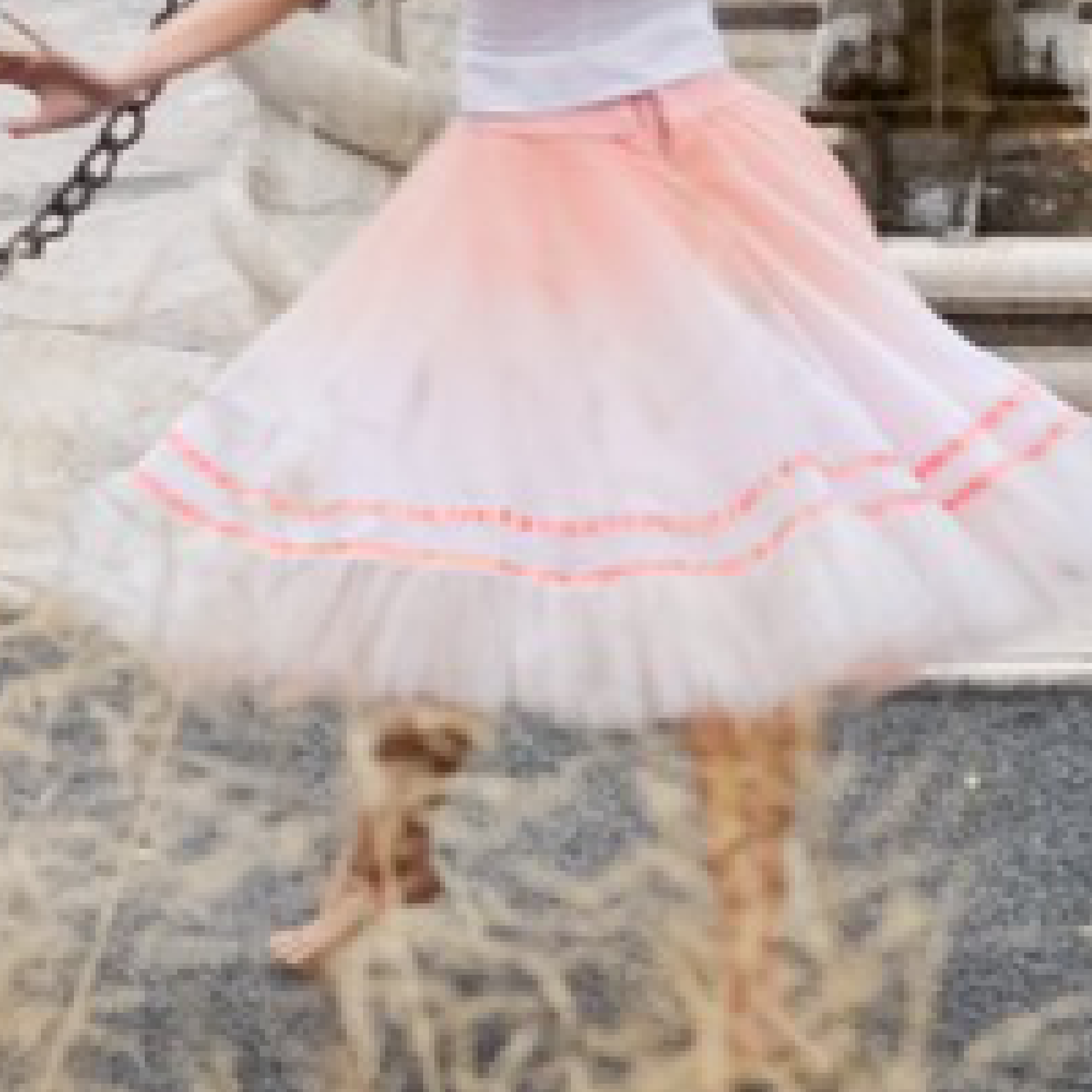}}\\[-0.9em]

    \vspace*{0.05cm}
    \caption{Qualitative examples from the DAVIS test set demonstrating the superiority and shortcomings of our approach. 
    }
	\label{fig:qualitative}
    \vspace*{-0.4cm}
\end{figure*}

\subsection{Knowledge Distillation}

To verify the effectiveness of the distillation loss, a `baseline' model was created by training the pruned model (obtained from the previous stage) on the ground-truth data using a Laplacian loss (the same training conditions as ST-MFNet). We note that, after 20 epochs, this baseline shows decreased average performance compared to the original model (shown in~\autoref{tbl:quantitative_comp} in Sec.~\ref{sec:results}).

In order to enhance the performance of the pruned model, previous approaches \cite{ding2021cdfi} have added new components to the model, at the cost of increased model size. In contrast here, we devise a knowledge distillation loss for VFI, where the pruned `student' network learns additional information from the pre-trained original ST-MFNet, the `teacher' network, without any increase in size.

Specifically, the loss function used to train the student network consists of two components. Firstly, the loss between the ground truth and the student model's prediction ($\mathcal{L}_{stud}$), and secondly, the loss between the student and the teacher's predictions ($\mathcal{L}_{dist}$). The total loss for the model is therefore
\begin{equation}
    \mathcal{L}_{total} =  \alpha \mathcal{L}_{stud}(I_{out}, I_{gt}) + \mathcal{L}_{dist}(I_{pred}, I_{out}) ,
    \label{eqn:loss}
\end{equation}
where $I_{gt}$ represents the ground truth frame, $I_{out}$ represents the student model's output, and $I_{pred}$ represents the original model's prediction. Here, $\alpha$ is a hyper-parameter controlling the penalty from the ground-truth frames. The student network is trained from scratch using the loss function (Eqn.~\ref{eqn:loss}), enabling the pruned network to make use of the additional knowledge learned by the teacher network in prior training.

\section{Results and Discussion}
\label{sec:results}

\subsection{Experimental Setup}
\label{sec:results-setup}

\textbf{Implementation details.} The student model is trained for 20 epochs on the same training set as the pre-trained teacher model, i.e. approximately 92K frame quintuplets, from the Vimeo90K septuplet~\cite{xue2019video} and BVI-DVC~\cite{ma2020bvi} datasets. We use Laplacian pyramid loss as both $\mathcal{L}_{stud}$ and $\mathcal{L}_{dist}$, and AdaMax~\cite{kingma2014adam} with $\beta_1=0.9,\beta_2=0.999$. The hyper-parameter $\alpha$ in Eqn.~\ref{eqn:loss} is set to 0.1, based on empirical results.

\noindent \textbf{Evaluation.} The model has been evaluated on multiple commonly used test datasets, including UCF101, DAVIS, and SNU-Film. Two commonly used metrics, PSNR and SSIM~\cite{wang2004image}, are used to evaluate the frame interpolation performance.

\subsection{Alternative Loss Functions}

The baseline and Laplacian distilled models are compared in \autoref{tbl:quantitative_comp}, where it can be observed that the distillation process significantly improves model performance. To investigate potential further improvements, particularly in terms of perceptual quality, several GAN-based loss functions were tested as the distillation loss. Specifically, the following adversarial functions were evaluated: GAN~\cite{goodfellow2020generative}, ST-GAN~\cite{yang2016stationary}, and FIGAN~\cite{lee2020adacof}. These were trained for 10 epochs on the same training set, with the same hyperparameters as before, and scored respective PSNR values of 32.718, 32.716, and 32.803 on UCF101. Since they failed to improve on the effectiveness of the Laplacian distillation, they were not trained further.

\subsection{Qualitative Evaluation}

\noindent\textbf{Visual comparisons.}
Examples of frames interpolated using our model and ST-MFNet are shown in \autoref{fig:qualitative}. Under certain conditions, the new model's performance is observably closer to the ground truth than that of ST-MFNet, as evidenced by the first example, while for footage containing large movements, such as for the second case, the model's performance suffers slightly.

\subsection{Quantitative Evaluation}

\begin{table*}[t]
\caption{Quantitative comparison results (PSNR/SSIM) for ST-MFNet Mini and 13 otjher tested methods. In some cases, scores found using pre-trained models are provided, underlined, when they outperform their re-trained counterparts. For each column, the best result is bold in \textbf{\textcolor{red}{red}} and the second best is italicised in \textit{\textcolor{blue}{blue}}. The average runtime (RT) for interpolating a 480p DAVIS frame as well as the number of model parameters (\#P) for each method are also reported.}
\vspace{-5mm}
	\begin{center}
		\resizebox{\linewidth}{!}{
			\begin{tabular}{lcccccccc}
				\toprule
				& \multirow{2}[1]{*}{UCF101} & \multirow{2}[1]{*}{DAVIS} & \multicolumn{4}{c}{SNU-FILM} & \multirow{2}[2]{*}{\makecell{RT \\ (sec)}} & \multirow{2}[2]{*}{\makecell{\#P \\ (M)}} \\
				\cmidrule(l{5pt}r{5pt}){4-7}
				& & & Easy & Medium & Hard & Extreme & & \\
				\midrule
				SuperSloMo~\cite{jiang2018super}    & 32.547/0.968 & 26.523/0.866  & 36.255/0.984 & 33.802/0.973 & 29.519/0.930 & 24.770/0.855 & 0.107 & 39.61\\
				SepConv~\cite{niklaus2017video}     & 32.524/0.968 & 26.441/0.853 & 39.894/0.990 & 35.264/0.976 & 29.620/0.926 & 24.653/0.851 & \textit{\textcolor{blue}{0.062}} & 21.68\\
				DAIN~\cite{bao2019depth}            & \underline{32.524}/\underline{0.968} & 27.086/0.873 & 39.280/0.989 & 34.993/0.976 & 29.752/0.929 & 24.819/0.850 & 0.896 & 24.03\\
				BMBC~\cite{park2020bmbc}            & \underline{32.729}/\underline{0.969} & 26.835/0.869 & 39.809/0.990 & 35.437/0.978 & 29.942/0.933 & 24.715/0.856 & 1.425 & 11.01\\
				AdaCoF~\cite{lee2020adacof}         & \underline{32.610}/\underline{0.968} & 26.445/0.854 & 39.912/0.990 & 35.269/0.977 & 29.723/0.928 & 24.656/0.851 & \textbf{\textcolor{red}{0.051}} & 21.84\\
				CDFI~\cite{ding2021cdfi}            & \underline{32.653}/\underline{0.968} & 26.471/0.857 & 39.881/0.990 & 35.224/0.977 & 29.660/0.929 & 24.645/0.854 & 0.321 & \textit{\textcolor{blue}{4.98}}\\
				CAIN~\cite{choi2020channel}         & \underline{32.537}/\underline{0.968} & 26.477/0.857 & \underline{39.890}/\underline{0.990} & 35.630/0.978 & 29.998/0.931 & 25.060/0.857 & 0.071 & 42.78\\
				SoftSplat~\cite{niklaus2020softmax} & 32.835/0.969 & \textit{\textcolor{blue}{27.582}}/0.881 & 40.165/0.991 & 36.017/0.979 & 30.604/0.937 & 25.436/0.864 & 0.206 & 12.46\\
				EDSC~\cite{cheng2021multiple}       & \underline{32.677}/\underline{0.969} & 26.689/0.860 & 39.792/0.990 & 35.283/0.977 & 29.815/0.929 & 24.872/0.854 & 0.067 & 8.95\\
				XVFI~\cite{sim2021xvfi}             & 32.224/0.966 & 26.565/0.863 & 38.849/0.989 & 34.497/0.975 & 29.381/0.929 & 24.677/0.855 & 0.108 & 5.61\\
				QVI~\cite{xu2019quadratic}           & 32.668/0.967 & 27.483/\textit{\textcolor{blue}{0.883}} & 36.648/0.985 & 34.637/0.978 & 30.614/0.947 & 25.426/0.866 & 0.257 & 29.23\\
				FLAVR~\cite{kalluri2020flavr}       & \underline{\textbf{\textcolor{red}{33.389}}}/\underline{\textbf{\textcolor{red}{0.971}}} & 27.450/0.873 & 40.135/0.990 & 35.988/0.979 & 30.541/0.937 & 25.188/0.860 & 0.695 & 42.06\\
				ST-MFNet~\cite{danier2022st}        & \textit{\textcolor{blue}{33.384}}/\textit{\textcolor{blue}{0.970}} & \textbf{\textcolor{red}{28.287}}/\textbf{\textcolor{red}{0.895}} & 40.775/\textbf{\textcolor{red}{0.992}} & \textbf{\textcolor{red}{37.111}}/\textbf{\textcolor{red}{0.985}} & \textbf{\textcolor{red}{31.698}}/\textbf{\textcolor{red}{0.951}} & \textit{\textcolor{blue}{25.810}}/0.874 & 0.901 & 21.03\\
                \midrule
                ST-MFNet Mini (Baseline) & 33.013/0.968 & 26.013/0.848 & \textit{\textcolor{blue}{40.826}}/\textit{\textcolor{blue}{0.991}} & 35.858/0.980 & 30.724/0.946 & 25.726/\textit{\textcolor{blue}{0.891}} & 0.588 & \textbf{\textcolor{red}{1.82}}\\
                
                ST-MFNet Mini (KD Loss) & 33.070/0.969 & 26.335/0.855 & \textbf{\textcolor{red}{40.982}}/\textbf{\textcolor{red}{0.992}} & \textit{\textcolor{blue}{36.127}}/\textit{\textcolor{blue}{0.982}} & \textit{\textcolor{blue}{30.914}}/\textit{\textcolor{blue}{0.950}} & \textbf{\textcolor{red}{25.877}}/\textbf{\textcolor{red}{0.894}}  & 0.588 & \textbf{\textcolor{red}{1.82}}\\

				\bottomrule
		\end{tabular}}
		\label{tbl:quantitative_comp}
  		\vspace{-8mm}
	\end{center}
\end{table*}

The quantitative evaluation results are summarised in \autoref{tbl:quantitative_comp}. It is noted that the KD-trained compressed model (ST-MFNet Mini) outperforms it's teacher on two databases: SNU-FILM Easy and Extreme, while its performance on databases with large motions are not as good compared to the state of the art. This may be explained by the fact that the BLFNet component of the model was removed, which was previously  shown to improve the model's ability to handle large motions~\cite{danier2022st}. 

\begin{figure}
    \centering
    \includegraphics[width=.95\linewidth]{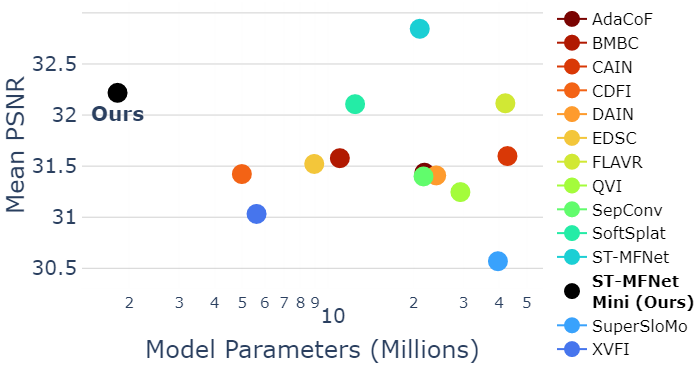}   
    \caption{Mean PSNR scores from the models presented in~\autoref{tbl:quantitative_comp}. On average, our method outperforms all models except its teacher, at a greatly reduced size.}
    \label{fig:model_comparison}
    \vspace{-4mm}
\end{figure}

\autoref{fig:model_comparison} plots the mean PSNR values of all the tested VFI approaches against their model parameters. It can be observed that STMFNet Mini achieves an excellent trade off between performance and model size, with the lowest model size but second best overall performance. The model with the second fewest model parameters, CDFI, is nearly three times as large. 

However, while the compressed model outperforms for its size, it does not meet the same expectations for its runtime. Despite being 35\% faster than the original ST-MFNet, the model is still significantly slower than other, larger models. This is most likely due to the upscaling component of ST-MFNet (3D CNN), which requires lengthy computation time; it will be a focus of our future work.

\section{Conclusion}
\label{sec:conclusion}

A workflow for frame interpolation model compression is presented which utilises model pruning to determine a reduced model architecture and knowledge distillation to improve its performance. When applied to an advanced VFI approach, ST-MFNet; the resulting low complexity model (ST-MFNet Mini) only requires 9\% of the original model's parameters to achieve competitive interpolation performance compared to the state-of-the-art. Future work should focus on generalising this workflow to other VFI approaches, employing more effective knowledge distillation implementations (e.g., feature based) during training in the second stage and on approaches that significantly reduce model runtime.

\small
\bibliographystyle{IEEEtran}
\bibliography{bibliography.bib}

\begin{thebibliography}{10}
\providecommand{\url}[1]{#1}
\csname url@samestyle\endcsname
\providecommand{\newblock}{\relax}
\providecommand{\bibinfo}[2]{#2}
\providecommand{\BIBentrySTDinterwordspacing}{\spaceskip=0pt\relax}
\providecommand{\BIBentryALTinterwordstretchfactor}{4}
\providecommand{\BIBentryALTinterwordspacing}{\spaceskip=\fontdimen2\font plus
\BIBentryALTinterwordstretchfactor\fontdimen3\font minus
  \fontdimen4\font\relax}
\providecommand{\BIBforeignlanguage}[2]{{%
\expandafter\ifx\csname l@#1\endcsname\relax
\typeout{** WARNING: IEEEtran.bst: No hyphenation pattern has been}%
\typeout{** loaded for the language `#1'. Using the pattern for}%
\typeout{** the default language instead.}%
\else
\language=\csname l@#1\endcsname
\fi
#2}}
\providecommand{\BIBdecl}{\relax}
\BIBdecl

\bibitem{dong2022video}
J.~Dong, K.~Ota, and M.~Dong, ``{Video Frame Interpolation: A Comprehensive
  Survey},'' \emph{ACM Transactions on Multimedia Computing, Communications and
  Applications}, 2022.

\bibitem{danier2022st}
D.~Danier, F.~Zhang, and D.~Bull, ``{ST-MFNet}: A spatio-temporal multi-flow
  network for frame interpolation,'' in \emph{Proceedings of the IEEE/CVF
  Conference on Computer Vision and Pattern Recognition}, 2022, pp. 3521--3531.

\bibitem{jiang2018super}
H.~Jiang, D.~Sun, V.~Jampani, M.-H. Yang, E.~Learned-Miller, and J.~Kautz,
  ``Super slomo: High quality estimation of multiple intermediate frames for
  video interpolation,'' in \emph{Proceedings of the IEEE Conference on
  Computer Vision and Pattern Recognition}, 2018, pp. 9000--9008.

\bibitem{xu2019quadratic}
X.~Xu, L.~Siyao, W.~Sun, Q.~Yin, and M.-H. Yang, ``Quadratic video
  interpolation,'' in \emph{NeurIPS}, 2019.

\bibitem{niklaus2020softmax}
S.~Niklaus and F.~Liu, ``Softmax splatting for video frame interpolation,'' in
  \emph{Proceedings of the IEEE/CVF Conference on Computer Vision and Pattern
  Recognition}, 2020, pp. 5437--5446.

\bibitem{niklaus2017video}
S.~Niklaus, L.~Mai, and F.~Liu, ``Video frame interpolation via adaptive
  separable convolution,'' in \emph{Proceedings of the IEEE International
  Conference on Computer Vision}, 2017, pp. 261--270.

\bibitem{lee2020adacof}
H.~Lee, T.~Kim, T.-y. Chung, D.~Pak, Y.~Ban, and S.~Lee, ``Adacof: Adaptive
  collaboration of flows for video frame interpolation,'' in \emph{Proceedings
  of the IEEE/CVF Conference on Computer Vision and Pattern Recognition}, 2020,
  pp. 5316--5325.

\bibitem{choi2020channel}
M.~Choi, H.~Kim, B.~Han, N.~Xu, and K.~M. Lee, ``Channel attention is all you
  need for video frame interpolation,'' in \emph{Proceedings of the AAAI
  Conference on Artificial Intelligence}, vol.~34, 2020, pp. 10\,663--10\,671.

\bibitem{reda2022film}
F.~Reda, J.~Kontkanen, E.~Tabellion, D.~Sun, C.~Pantofaru, and B.~Curless,
  ``{FILM}: Frame interpolation for large motion,'' in \emph{Computer
  Vision--ECCV 2022: 17th European Conference, Tel Aviv, Israel, October
  23--27, 2022, Proceedings, Part VII}.\hskip 1em plus 0.5em minus 0.4em\relax
  Springer, 2022, pp. 250--266.

\bibitem{kong2022ifrnet}
L.~Kong, B.~Jiang, D.~Luo, W.~Chu, X.~Huang, Y.~Tai, C.~Wang, and J.~Yang,
  ``Ifrnet: Intermediate feature refine network for efficient frame
  interpolation,'' in \emph{Proceedings of the IEEE/CVF Conference on Computer
  Vision and Pattern Recognition}, 2022, pp. 1969--1978.

\bibitem{bao2019depth}
W.~Bao, W.-S. Lai, C.~Ma, X.~Zhang, Z.~Gao, and M.-H. Yang, ``Depth-aware video
  frame interpolation,'' in \emph{Proceedings of the IEEE/CVF Conference on
  Computer Vision and Pattern Recognition}, 2019, pp. 3703--3712.

\bibitem{bao2019memc}
W.~Bao, W.-S. Lai, X.~Zhang, Z.~Gao, and M.-H. Yang, ``Memc-net: Motion
  estimation and motion compensation driven neural network for video
  interpolation and enhancement,'' \emph{IEEE transactions on pattern analysis
  and machine intelligence}, vol.~43, no.~3, pp. 933--948, 2019.

\bibitem{gui2020featureflow}
S.~Gui, C.~Wang, Q.~Chen, and D.~Tao, ``Featureflow: Robust video interpolation
  via structure-to-texture generation,'' in \emph{Proceedings of the IEEE/CVF
  Conference on Computer Vision and Pattern Recognition}, 2020, pp.
  14\,004--14\,013.

\bibitem{kalluri2020flavr}
T.~Kalluri, D.~Pathak, M.~Chandraker, and D.~Tran, ``Flavr: Flow-agnostic video
  representations for fast frame interpolation,'' \emph{arXiv preprint
  arXiv:2012.08512}, 2020.

\bibitem{danier2022enhancing}
D.~Danier, F.~Zhang, and D.~Bull, ``Enhancing deformable convolution based
  video frame interpolation with coarse-to-fine 3d cnn,'' in \emph{2022 IEEE
  International Conference on Image Processing (ICIP)}.\hskip 1em plus 0.5em
  minus 0.4em\relax IEEE, 2022, pp. 1396--1400.

\bibitem{lu2022video}
L.~Lu, R.~Wu, H.~Lin, J.~Lu, and J.~Jia, ``Video frame interpolation with
  transformer,'' in \emph{Proceedings of the IEEE/CVF Conference on Computer
  Vision and Pattern Recognition}, 2022, pp. 3532--3542.

\bibitem{shi2022video}
Z.~Shi, X.~Xu, X.~Liu, J.~Chen, and M.-H. Yang, ``Video frame interpolation
  transformer,'' in \emph{Proceedings of the IEEE/CVF Conference on Computer
  Vision and Pattern Recognition}, 2022, pp. 17\,482--17\,491.

\bibitem{lacoste2019quantifying}
A.~Lacoste, A.~Luccioni, V.~Schmidt, and T.~Dandres, ``Quantifying the carbon
  emissions of machine learning,'' \emph{arXiv preprint arXiv:1910.09700},
  2019.

\bibitem{reed1993pruning}
R.~Reed, ``Pruning algorithms-a survey,'' \emph{IEEE transactions on Neural
  Networks}, vol.~4, no.~5, pp. 740--747, 1993.

\bibitem{hinton2015distilling}
G.~Hinton, O.~Vinyals, and J.~Dean, ``Distilling the knowledge in a neural
  network,'' \emph{arXiv preprint arXiv:1503.02531}, 2015.

\bibitem{ding2021cdfi}
T.~Ding, L.~Liang, Z.~Zhu, and I.~Zharkov, ``{CDFI}: Compression-driven network
  design for frame interpolation,'' in \emph{Proceedings of the IEEE/CVF
  Conference on Computer Vision and Pattern Recognition}, 2021, pp. 8001--8011.

\bibitem{chen2021orthant}
T.~Chen, T.~Ding, B.~Ji, G.~Wang, Y.~Shi, J.~Tian, S.~Yi, X.~Tu, and Z.~Zhu,
  ``Orthant based proximal stochastic gradient method for $l_1$-regularized
  optimization,'' in \emph{Machine Learning and Knowledge Discovery in
  Databases: European Conference, ECML PKDD 2020, Ghent, Belgium, September
  14--18, 2020, Proceedings, Part III}.\hskip 1em plus 0.5em minus 0.4em\relax
  Springer, 2021, pp. 57--73.

\bibitem{fourure2017residual}
D.~Fourure, R.~Emonet, E.~Fromont, D.~Muselet, A.~Tremeau, and C.~Wolf,
  ``Residual conv-deconv grid network for semantic segmentation,'' \emph{arXiv
  preprint arXiv:1707.07958}, 2017.

\bibitem{chen2020neural}
T.~Chen, B.~Ji, Y.~Shi, T.~Ding, B.~Fang, S.~Yi, and X.~Tu, ``Neural network
  compression via sparse optimization,'' \emph{arXiv preprint
  arXiv:2011.04868}, 2020.

\bibitem{xue2019video}
T.~Xue, B.~Chen, J.~Wu, D.~Wei, and W.~T. Freeman, ``Video enhancement with
  task-oriented flow,'' \emph{International Journal of Computer Vision}, vol.
  127, no.~8, pp. 1106--1125, 2019.

\bibitem{ma2020bvi}
D.~Ma, F.~Zhang, and D.~Bull, ``{BVI-DVC}: A training database for deep video
  compression,'' \emph{IEEE Transactions on Multimedia}, pp. 1--1, 2021.

\bibitem{kingma2014adam}
D.~P. Kingma and J.~Ba, ``Adam: A method for stochastic optimization,''
  \emph{arXiv preprint arXiv:1412.6980}, 2014.

\bibitem{wang2004image}
Z.~Wang, A.~C. Bovik, H.~R. Sheikh, and E.~P. Simoncelli, ``Image quality
  assessment: from error visibility to structural similarity,'' \emph{IEEE
  transactions on image processing}, vol.~13, no.~4, pp. 600--612, 2004.

\bibitem{goodfellow2020generative}
I.~Goodfellow, J.~Pouget-Abadie, M.~Mirza, B.~Xu, D.~Warde-Farley, S.~Ozair,
  A.~Courville, and Y.~Bengio, ``Generative adversarial networks,''
  \emph{Communications of the ACM}, vol.~63, no.~11, pp. 139--144, 2020.

\bibitem{yang2016stationary}
F.~Yang, G.-S. Xia, L.~Zhang, and X.~Huang, ``Stationary dynamic texture
  synthesis using convolutional neural networks,'' in \emph{2016 IEEE 13th
  International Conference on Signal Processing (ICSP)}.\hskip 1em plus 0.5em
  minus 0.4em\relax IEEE, 2016, pp. 1135--1139.

\bibitem{park2020bmbc}
J.~Park, K.~Ko, C.~Lee, and C.-S. Kim, ``{BMBC}: Bilateral motion estimation
  with bilateral cost volume for video interpolation,'' in \emph{Computer
  Vision--ECCV 2020: 16th European Conference, Glasgow, UK, August 23--28,
  2020, Proceedings, Part XIV 16}.\hskip 1em plus 0.5em minus 0.4em\relax
  Springer, 2020, pp. 109--125.

\bibitem{cheng2021multiple}
X.~Cheng and Z.~Chen, ``Multiple video frame interpolation via enhanced
  deformable separable convolution,'' \emph{IEEE Transactions on Pattern
  Analysis and Machine Intelligence}, 2021.

\bibitem{sim2021xvfi}
H.~Sim, J.~Oh, and M.~Kim, ``{XVFI}: extreme video frame interpolation,'' in
  \emph{Proceedings of the IEEE International Conference on Computer Vision
  (ICCV)}, 2021.

\end{thebibliography}

\end{document}